\documentclass[conference]{IEEEtran}
\IEEEoverridecommandlockouts
\usepackage{cite}
\usepackage{amsmath,amssymb,amsfonts}
\usepackage{algorithmic}
\usepackage{graphicx}
\usepackage{textcomp}
\usepackage[numbers]{natbib}
\usepackage{fancyhdr} 
\usepackage{array}    
\usepackage{xcolor}
\usepackage{booktabs} 
\usepackage{makecell} 
\usepackage[normalem]{ulem} 
\def\BibTeX{{\rm B\kern-.05em{\sc i\kern-.025em b}\kern-.08em
    T\kern-.1667em\lower.7ex\hbox{E}\kern-.125emX}}

\begin{document}

\title{SFE-Net: Harnessing Biological Principles of Differential Gene Expression for Improved Feature Selection in Deep Learning Networks\\
}

\author{\IEEEauthorblockN{1\textsuperscript{st} Yuqi Li}
\IEEEauthorblockA{\textit{Dept. of Large Language Model} \\
\textit{Qifu Technology}\\
Shanghai, China \\
liyuqi1-jk@360shuke.com}

\and
\IEEEauthorblockN{2\textsuperscript{nd} Yuanzhong Zheng}
\IEEEauthorblockA{\textit{Dept. of Large Language Model} \\
\textit{Qifu Technology}\\
Shanghai, China \\
zhengyuanzhong-jk@360shuke.com}
\and
\IEEEauthorblockN{3\textsuperscript{rd} Yaoxuan Wang}
\IEEEauthorblockA{\textit{Dept. of Large Language Model} \\
\textit{Qifu Technology}\\
Shanghai, China \\
wangyaoxuan-jk@360shuke.com}
\and
\IEEEauthorblockN{\hspace{20mm}4\textsuperscript{th} Jianjun Yin*}
\IEEEauthorblockA{\hspace{20mm}\textit{School Of Information Science And Technology} \\
\textit{\hspace{20mm} Fudan University}\\
\hspace{20mm} Shanghai, China \\
\hspace{20mm}yinjianjun@fudan.edu.cn}
\and
\IEEEauthorblockN{5\textsuperscript{th} Haojun Fei*}
\IEEEauthorblockA{\textit{Dept. of Large Language Model} \\
\textit{Qifu Technology}\\
Shanghai, China \\
feihaojun-jk@360shuke.com}
}

\maketitle

\begin{abstract}
In the realm of DeepFake detection, the challenge of adapting to various synthesis methodologies such as Faceswap, Deepfakes, Face2Face, and NeuralTextures significantly impacts the performance of traditional machine learning models. These models often suffer from static feature representation, which struggles to perform consistently across diversely generated deepfake datasets. Inspired by the biological concept of differential gene expression, where gene activation is dynamically regulated in response to environmental stimuli, we introduce the Selective Feature Expression Network (SFE-Net). This innovative framework integrates selective feature activation principles into deep learning architectures, allowing the model to dynamically adjust feature priorities in response to varying deepfake generation techniques. SFE-Net employs a novel mechanism that selectively enhances critical features essential for accurately detecting forgeries, while reducing the impact of irrelevant or misleading cues akin to adaptive evolutionary processes in nature. Through rigorous testing on a range of deepfake datasets, SFE-Net not only surpasses existing static models in detecting sophisticated forgeries but also shows enhanced generalization capabilities in cross-dataset scenarios. Our approach significantly mitigates overfitting by maintaining a dynamic balance between feature exploration and exploitation, thus producing more robust and effective deepfake detection models. This bio-inspired strategy paves the way for developing adaptive deep learning systems that are finely tuned to address the nuanced challenges posed by the varied nature of digital forgeries in modern digital forensics.
\end{abstract}

\begin{IEEEkeywords}
DeepFake detection,SFE-Net,cross-dataset generalization,dynamic feature adjustment
\end{IEEEkeywords}

\section{Introduction}
\begin{figure}[htbp]
\centerline{
    \includegraphics[width=\linewidth, trim=2mm 85mm 100mm 10mm, clip]{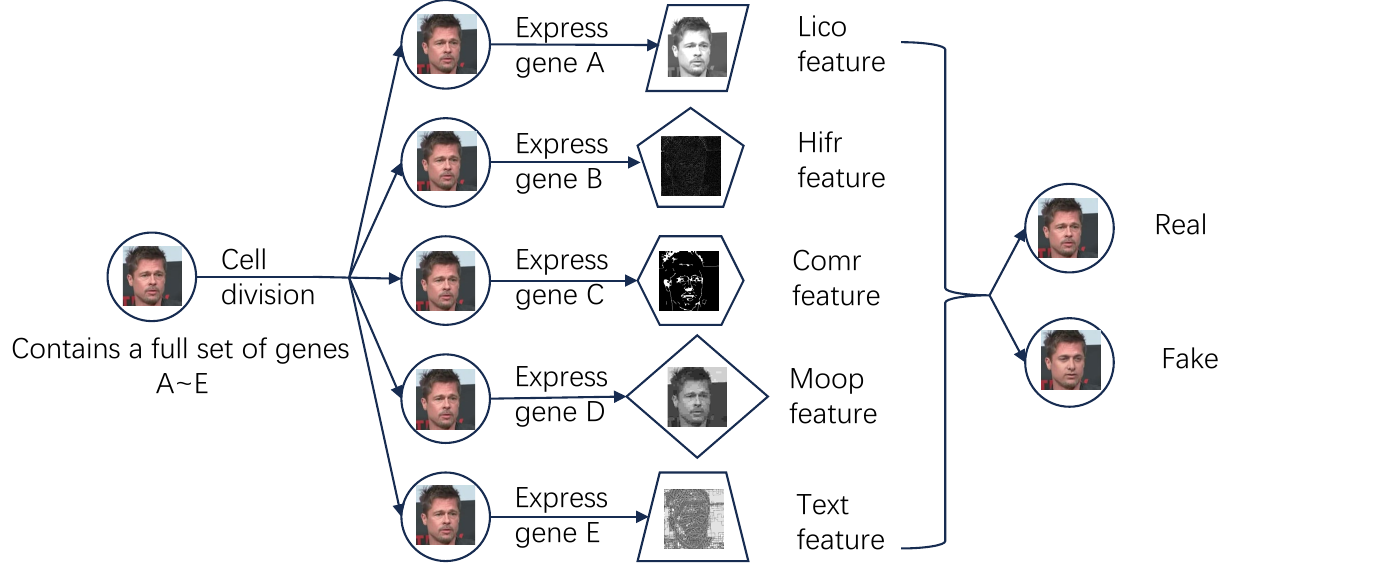}
}
\caption{Introducing feature-selective expression based on gene-selective expression: Lico--light consistency feature, Hifr-- high frequency feature, Comr--compression reconstruction feature, Moop--morphological operation feature, Text--texture feature}
\label{fig:SFE}
\end{figure}

In the domain of digital media, deepfake technology has emerged as a double-edged sword. On one hand, it facilitates innovative applications across entertainment, personalization, and education, exemplified by its capacity to generate hyper-realistic content seamlessly\cite{branwen2019gpt}\cite{westerlund2019emergence}\cite{vincent2018watch}\cite{suwajanakorn2017synthesizing}. On the other hand, the misuse of this technology poses substantial risks, undermining trust in media, infringing upon privacy, and enabling misinformation \cite{fallis2019fake}\cite{de2021distinct}\cite{vaccari2020deepfakes}.The evolution of deepfake technology, powered by advancements in generative adversarial networks (GANs) and deep learning, calls for equally sophisticated detection mechanisms to mitigate these threats.

Detection methods have traditionally utilized various strategies with distinct approaches and effectiveness. Simpler methods, such as Meso4\cite{afchar2018mesonet}, MesoInception\cite{afchar2018mesonet}, CNN-Aug\cite{CNN-Aug}, Xception\cite{FF++}, and EfficientNetB4\cite{tan2019efficientnet}, provide foundational insights into identifying basic manipulations but often falter with complex or subtle forgeries.

To overcome the shortcomings of these naive methods, specialized strategies have been developed. Spatial approaches like CapsuleNet\cite{nguyen2019capsule}, FWA (Face Warping Artifacts)\cite{FWA}, Face X-ray\cite{facex-ray}, FFD (Full Face Detection)\cite{FFD}, CORE\cite{ni2022core}, Recce\cite{Recce}, and UCF\cite{yan2023ucf} focus on visual inconsistencies at the pixel level, excelling at detecting irregular patterns like blending boundaries and unnatural lighting.

Frequency-based approaches analyze the frequency domain to identify anomalies, with techniques like F3Net\cite{f3net}, SPSL (Spatial-Phase Shallow Learning)\cite{spsl}, and SRM (Spectral Residual Method)\cite{SRM} being particularly effective against advanced forgeries that introduce subtle artifacts while maintaining visual plausibility. 

Building on this foundation, our research introduces the Selective Feature Expression Network (SFE-Net), inspired by the dynamic and selective response mechanisms observed in biological systems.We show in Fig. \ref{fig:SFE}. how to understand feature selective expression based on gene selective expression.  Just as organisms adaptively modulate their gene expression in response to environmental changes, SFE-Net dynamically adjusts its feature selection based on the characteristics of the deepfake input it analyzes. This approach, which draws parallels to gene selective expression, enhances the model’s adaptability and effectiveness against evolving deepfake techniques. By incorporating principles from natural evolutionary processes, SFE-Net not only targets specific types of deepfakes but also significantly improves its generalizability across different datasets and deepfake generation methods. This ability to maintain high performance in varied conditions positions our model at the forefront of digital forensics, pioneering robust and adaptive solutions for deepfake detection.

\section{Methodology}
\begin{figure}[htbp]
    \centering
    \includegraphics[width=\linewidth, trim=80mm 10mm 110mm 15mm, clip]{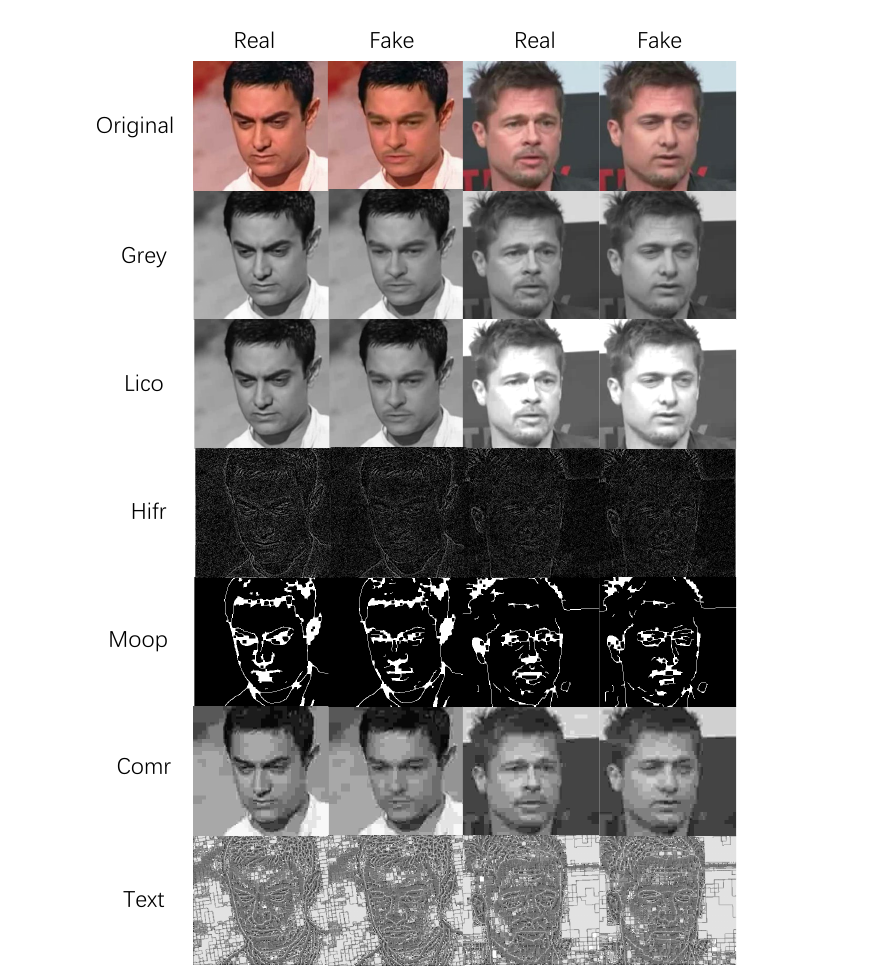}
    \caption{Visualizing the comparison of different feature extractions.}
    \label{fig:all_picture}
\end{figure}

In the pursuit of effective DeepFake detection, our methodology leverages a combination of traditional and advanced digital image processing techniques to identify anomalies that commonly arise in manipulated images. Below, we delineate the specific methods employed to detect various types of DeepFake technologies.
Additionally, Fig. \ref{fig:all_picture}.  displays the effects and comparisons of different feature extraction methods used in our analysis.

\subsection{FaceSwap Detection}
For detecting FaceSwap manipulations, we utilize morphological operations such as dilation and erosion, followed by opening and closing techniques, to examine and enhance the natural continuity of facial edges, allowing for the detection of unnatural edge formations typical in FaceSwap images. 

\begin{figure*}[htbp]
    \centering
    \includegraphics[width=\textwidth, height=0.35\textheight, keepaspectratio, trim=35mm 45mm 15mm 5mm, clip]{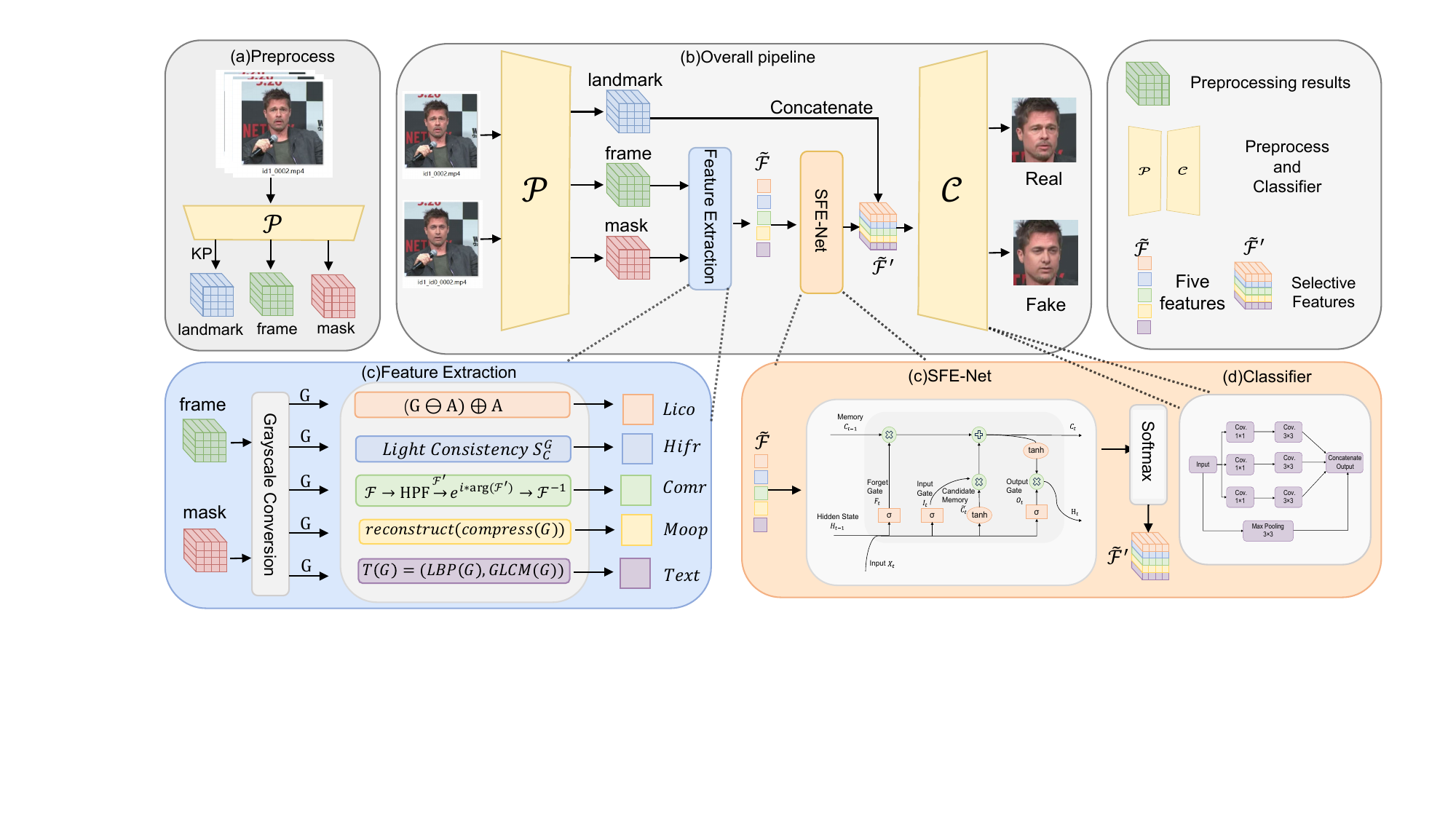}
    \caption{SFE-Net architecture}
    \label{fig:different feature}
\end{figure*}

\subsubsection*{Erosion}
Erosion of an image \( A \) by a structuring element \( B \) can be defined as:
\[
A \ominus B = \{z \mid B_z \subseteq A\}
\]
where \( B_z \) is the translation of \( B \) centered at point \( z \).

\subsubsection*{Dilation}
Dilation of an image \( A \) by a structuring element \( B \) can be defined as:
\[
A \oplus B = \{z \mid (\hat{B})_z \cap A \neq \emptyset\}
\]
where \( \hat{B} \) is the reflection of \( B \) about its origin.

\subsubsection*{Opening}
Opening of an image is an erosion followed by a dilation:
\[
(A \ominus B) \oplus B
\]
This operation is useful for removing small noise while preserving the shape and size of larger objects in the image.







Building upon our method, we also evaluate lighting consistency across facial regions using algorithms designed to detect abnormal lighting patterns, which are crucial for identifying digital manipulations. For an image tensor \( I \) with dimensions \( (B, H, W, C) \), where \( B \) is the batch size, \( H \) the height, \( W \) the width, and \( C \) the number of channels, the lighting consistency score \( S \) for each image can be calculated as follows:

\[
S = \text{var}(I_c) = \frac{1}{HW} \sum_{i=1}^{H} \sum_{j=1}^{W} (I_{cij} - \mu_c)^2
\]

where \( \mu_c \), the mean intensity of channel \( c \), is given by:

\[
\mu_c = \frac{1}{HW} \sum_{i=1}^{H} \sum_{j=1}^{W} I_{cij}
\]

This variance quantifies the dispersion of lighting across the image channel, with higher values indicating more inconsistency, thus aiding in the detection of forgeries.

\subsection{DeepFakes Detection}

Manipulation detection utilizes the Fourier Transform:
\[
F(u, v) = \mathcal{F}\{f(x, y)\}, \quad F'(u, v) = F(u, v) \cdot H(u, v)
\]
where \( \mathcal{F} \) denotes the Fourier Transform of the spatial domain image \( f(x, y) \). The high-pass filter \( H(u, v) \) is defined by:
\[
H(u, v) = 
\begin{cases} 
0 & \text{if } (u, v) \text{ is central} \\
1 & \text{otherwise}
\end{cases}
\]

\subsubsection*{Phase Extraction and Image Reconstruction}
The phase spectrum \( \phi(u, v) \) is extracted for reconstruction:
\begin{align*}
\phi(u, v) &= \arg(F'(u, v)), \\
G(u, v) &= e^{i\phi(u, v)}, \\
g(x, y) &= \mathcal{F}^{-1}\{G(u, v)\}
\end{align*}
This compact approach emphasizes phase information to effectively detect manipulations.

\subsection{Face2Face Detection}
\begin{table*}[htbp]
\centering
\caption{Comparison of DeepFake Detection Methods}
\label{tab:deepfake_methods}
\setlength{\tabcolsep}{10pt} 
\renewcommand{\arraystretch}{1.3} 
\begin{tabular}{@{}ccccccccc@{}}
\toprule
\textbf{Method} & \textbf{Detector} & \textbf{Backbone} & \textbf{CDF-v1} & \textbf{CDF-v2} & \textbf{DFD} & \textbf{DFDC} & \textbf{DFDCP} & \textbf{Avg.} \\
\midrule
Naive & Meso4 \cite{afchar2018mesonet} & MesoNet & 0.736 & 0.609 & 0.548 & 0.556 & 0.599 & 0.610 \\
Naive & MesoIncep \cite{afchar2018mesonet} & MesoNet & 0.737 & 0.697 & 0.607 & 0.623 & 0.756 & 0.684 \\
Naive & CNN-Aug \cite{CNN-Aug} & ResNet & 0.742 & 0.703 & 0.646 & 0.636 & 0.617 & 0.669 \\
Naive & Xception \cite{FF++}  & Xception & 0.779 & 0.737 & \uline{0.816} & 0.708 & 0.737 & 0.755 \\
Naive & EfficientB4 \cite{tan2019efficientnet}  & EfficientNet & 0.791 & 0.749 & 0.815 & 0.696 & 0.728 & 0.756 \\
\midrule 
Spatial & CapsuleNet \cite{nguyen2019capsule}  & Capsule & 0.791 & 0.747 & 0.684 & 0.647 & 0.657 & 0.705 \\
Spatial & FWA \cite{FWA}  & Xception & 0.790 & 0.668 & 0.740 & 0.613 & 0.638 & 0.690 \\
Spatial & Face X-ray \cite{facex-ray}  & HRNet & 0.709 & 0.679 & 0.766 & 0.633 & 0.694 & 0.696 \\
Spatial & FFD \cite{FFD} & Xception & 0.784 & 0.744 & 0.802 & 0.703 & 0.743 & 0.755 \\
Spatial & CORE \cite{ni2022core}  & Xception & 0.780 & 0.743 & 0.802 & 0.705 & 0.734 & 0.753 \\
Spatial & Recce \cite{Recce} & Designed & 0.768 & 0.732 & 0.812 & \uline{0.713} & 0.734 & 0.752 \\
Spatial & UCF \cite{yan2023ucf}  & Xception & 0.779 & 0.753 & 0.807 & \textbf{0.719} & \uline{0.759} & 0.763 \\
\midrule 
Frequency & F3Net \cite{f3net}& Xception & 0.777 & 0.735 & 0.798 & 0.702 & 0.735 & 0.749 \\
Frequency & SPSL \cite{spsl} & Xception & \uline{0.815} & \uline{0.765} & 0.812 & 0.704 & 0.741 & \uline{0.767} \\
Frequency & SRM \cite{SRM} & Xception & 0.793 & 0.755 & 0.812 & 0.700 & 0.741 & 0.760 \\
\midrule 
Frequency(Ours) & SFE-Net & Xception & \textbf{0.866}  & \textbf{0.798}  & \textbf{0.840}  & 0.709  & \textbf{0.760}  & \textbf{0.795} \\
\bottomrule
\end{tabular}
\end{table*}

Image Compression and Reconstruction: We analyze the effects of image compression and subsequent reconstruction to detect unnatural artifacts and inconsistencies in expressions. This technique is based on the premise that Face2Face manipulations may not hold up well under transformations that include compression, potentially revealing discrepancies that are not easily visible otherwise.

\subsection{NeuralTextures Detection}
Texture Analysis: We conduct a detailed analysis of the texture distribution within the image, using methods like Local Binary Patterns (LBP) and Gray Level Co-occurrence Matrix (GLCM). The goal is to identify any unnatural changes in texture that could suggest tampering.

\section{Experiment}
\subsection{Dataset}
In this study, we enhance DeepFake video detection algorithms using multiple datasets, including the extensive "FaceForensics++"\cite{FF++} for training, and others like Celeb-DF-v1\cite{li2020celeb}, Celeb-DF-v2\cite{li2020celeb}, DFDCP\cite{dfdcp} (DeepFake Detection Challenge Preview), DFDC\cite{dfdc} (DeepFake Detection Challenge),and DeepFakeDetection\cite{deepfakedetection2021} for robust testing across various scenarios. This approach helps refine and validate our models to effectively counteract a wide spectrum of DeepFake videos.

\subsection{Evaluation Metrics}
Typically, we evaluate our method using the frame-level Area Under Curve (AUC) metric, which assesses the model's ability to differentiate between classes at various thresholds, offering a comprehensive performance view. Additionally, we report video-level AUC, Average Precision (AP), and Equal Error Rate (EER) to provide further insights into the precision-recall balance and the point where false accept and reject rates are equal.

\subsection{Architecture}
The overall architecture of our proposed model is divided into four parts, as shown in Fig. \ref{fig:different feature}.
\begin{itemize}
    \item \textbf{Preprocess}: 
    The original input video is preprocessed to obtain frame-level RGB images,mask images and landmarks associated with key points.
    \item \textbf{Overall Pipeline}: 
    The input frame and landmark are both images. These RGB images are input into the feature extraction , where five different features are extracted to detect pseudo-images generated by DeepFake technology. In the SFE-Net, each of these features is fed into corresponding LSTM and softmax layers to further extract discontinuities between adjacent frames.
    \item \textbf{Classifier}
    Subsequently, these extracted features, along with the landmark features, are input into the classifier to discern the authenticity of the videos.
    \item \textbf{Output}
    The final output of the classifier indicates whether the video is 'Real' or 'Fake', effectively identifying and countering DeepFake manipulations by analyzing image-based features and those articulated through the SFE-Net.
\end{itemize}

\subsection{Ablation Experiment}

Based on the method we proposed, we give the ablation experiment results, where text~moop represent the five extracted features. The entire experiment is consistent with the evaluation process in the previous article. Finally, the average results after testing on 5 data sets are given.

\begin{table}[ht]
\centering
\caption{Ablation experiment results}
\label{my-label}
\begin{tabular}{@{}cccccc|c@{}}  
\toprule
Text & Comr & Hifr & Lico & Moop & SFE-Net & Avg. \\ \midrule
$\checkmark$ &            &             &            &            &         & 0.726 \\
             & $\checkmark$&            &            &            &         & 0.763 \\
            &              &$\checkmark$ &            &            &         & 0.751 \\
 &  &  & $\checkmark$ & &         & 0.745 \\
 &  &  &  & $\checkmark$ & & 0.767 \\
$\checkmark$ & $\checkmark$ &   $\checkmark$ & $\checkmark$ & $\checkmark$ &         & 0.771 \\
$\checkmark$ & $\checkmark$ & $\checkmark$ & $\checkmark$ & $\checkmark$ & $\checkmark$ & 0.795 \\ \bottomrule
\end{tabular}
\end{table}

\subsection{Performance}
Table \ref{tab:deepfake_methods} demonstrates the performance of various DeepFake detection methods, categorized as ``Naive,'' ``Spatial,'' and ``Frequency.'' These methods are evaluated using the Area Under Curve (AUC) metric across multiple datasets such as CDF-v1, CDF-v2, DFD, DFDC, and DFDCP. All detectors are trained on FF++ c23 \cite{FF++} and evaluated on the aforementioned datasets. The best results in each category are highlighted in bold, with the second-best results underlined.

Our proposed method, using the SFE-Net under the ``Frequency'' category, consistently outperforms others, achieving high AUC scores such as 0.866 on CDF-v1 and 0.795 as an impressive average across all datasets. This highlights the robustness and efficiency of SFE-Net in detecting DeepFake videos.The evaluation of other models here uses the weight files provided by \cite{DeepfakeBench_YAN_NEURIPS2023} .

While SFE-Net generally outperforms other models, its lower AUC of 0.709 on the DFDC dataset indicates areas for improvement. The complexity and variety of the DFDC's forgery techniques and visual conditions suggest that enhancements in feature extraction and model training are necessary. Addressing these issues is essential for advancing our DeepFake detection capabilities and maintaining effectiveness in real-world scenarios.

\section{Conclution}



We introduced the Selective Feature Expression Network (SFE-Net), a DeepFake detection model inspired by biological processes, capable of dynamically adjusting feature sensitivities. SFE-Net achieves superior detection performance and cross-dataset generalization on datasets like FaceForensics++ and Celeb-DF. Despite limitations such as performance variance on complex datasets,relatively high computational costs, and preprocessing dependence, this work enhances digital media integrity and lays the groundwork for adaptive deep learning models to combat evolving digital threats.


\clearpage

\vspace{12pt}

\end{document}